\documentstyle[preprint,aps,prd]{revtex}

\begin{document}
\draft
\title{One-Parameter Squeezed Gaussian States
of Time-Dependent Harmonic Oscillator and Selection
Rule for Vacuum States}

\author{Jung Kon Kim and Sang Pyo Kim}

\address{Department of Physics,
Kunsan National University,
Kunsan 573-701, Korea}
\date{\today}

\maketitle
\begin{abstract}
By using the invariant method we find one-parameter
squeezed Gaussian states for both time-independent
and time-dependent oscillators.
The squeezing parameter is expressed in terms of
energy expectation value for time-independent case
and represents the degree of mixing positive and
negative frequency solutions for time-dependent case.
A {\it minimum uncertainty proposal} is advanced to select uniquely
vacuum states at each moment of time.
We show that the Gaussian states with minimum uncertainty
coincide with the true vacuum state for time-independent
oscillator and the Bunch-Davies vacuum for a massive
scalar field in a de Sitter spacetime.
\end{abstract}

\section{Introduction}

Harmonic oscillators have played many important roles in
quantum physics,  partly because they are exactly solvable
quantum mechanically and partly because any
system around an equilibrium can be approximated as a harmonic
oscillator system. As a non-stationary system, a time-dependent
quantum harmonic oscillator can also be exactly
solved. One encounters  typical time-dependent harmonic oscillators
in a system of harmonic oscillators interacting with
an environment or evolving in an expanding
universe. In the former case, the harmonic oscillator
system depends on time through parametric couplings to
the environment. In the latter case, for instance, a massive
scalar field, as a collection of harmonic oscillators
when appropriately decomposed into modes, gains time-dependence
from a time-dependent spacetime background.
As a method to find the exact quantum states of a time-dependent
harmonic oscillator, Lewis and Riesenfeld
\cite{lewis,lewis2} have introduced an invariant, quadratic in momentum
and position, which satisfies the quantum Liouville-Neumann equation.
The exact quantum states are given by the eigenstates of this
invariant up to some time-dependent phase factors.
Since then there have been many variants and
applications of the invariants  and researches on the nature of the squeezed
states of the vacuum states[3-29].

In this paper, first we circumvent technically the task of solving a
time-dependent nonlinear auxiliary equation in terms of which the quadratic
invariant was expressed by Lewis and Riesenfeld \cite{lewis,lewis2},
by finding a pair of first order invariants in terms of a complex
solution to the classical equation and showing that the amplitude of the complex
solution satisfies the auxiliary equation.
By using the invariant method we find one-parameter squeezed Gaussian
states which are symmetric about the origin. The squeezing parameter
is determined by the energy expectation value for a time-independent
oscillator and represents the degree of mixing positive and negative
frequency solutions for a time-dependent oscillator.
Second, we propose the {\it minimum uncertainty} as a rule to select
uniquely the vacuum states for either time-independent and
time-dependent oscillators. The Gaussian states with the minimum
uncertainty have also the minimum energy expectation value
at every moment of time. The Gaussian states with minimum uncertainty
coincide with the true vacuum state of time-independent oscillator
and with the Bunch-Davies vacuum state for a minimal massive
scalar field in a de Sitter spacetime.

The organization of this paper is as follows. In Sec. II
we introduce a pair of first order invariants equivalent to the
original quadratic invariant by Lewis and Riesenfeld
and find one-parameter squeezed Gaussian states.
In Sec. III we study the minimum uncertainty as a selection
rule for vacuum states both for time-independent oscillator
and for time-dependent oscillator.

\section{One-Parameter Squeezed Gaussian States}

First, we show the equivalence between the quadratic invariant introduced
by Lewis and Riesenfeld \cite{lewis,lewis2} and a pair of
first order invariants.
Lewis and Riesenfeld
let us solve the time-dependent Schr\"{o}dinger equation
in the Schr\"{o}dinger-picture ($\hbar =1$)
\begin{equation}
i \frac{\partial}{\partial t} \Psi (q, t)
= \hat{H} (t) \Psi (q, t),
\label{sch eq}
\end{equation}
for a time-dependent harmonic oscillator of the form
\begin{equation}
\hat{H} = \frac{1}{2m_0} \hat{p}^2 + \frac{m_0 \omega^2 (t) }{2} \hat{q}^2.
\label{har osc2}
\end{equation}
Lewis and Riesenfeld introduced the invariant operator quadratic
in position and momentum
\begin{equation}
\hat{I} (t) = \frac{1}{2m_0} \Bigl[ (\xi \hat{p} - \dot{\xi} \hat{q})^2 +
\frac{1}{\xi^2} \hat{q}^2 \Bigr],
\label{inv}
\end{equation}
that satisfies the quantum Liouville-Neumann equation
\begin{equation}
i \frac{\partial}{\partial t} \hat{I} +
\Bigl[\hat{I}, \hat{H} \Bigr] = 0.
\end{equation}
Then $\xi$ satisfies the auxiliary equation
\begin{equation}
\ddot{\xi} + \omega^2 (t) \xi = \frac{1}{\xi^3}.
\label{aux}
\end{equation}

Instead of the quadratic invariant (\ref{inv}), let
us consider a pair of the first order invariants \cite{kim1}
\begin{eqnarray}
\hat{A} (t) &=& i \Bigl(u^* (t) \hat{p}
- m_0 \dot{u}^* (t) \hat{q} \Bigr),
\nonumber\\
\hat{A}^{\dagger} (t) &=& -i \Bigl( u(t) \hat{p}
- m_0 \dot{u} (t) \hat{q} \Bigr).
\label{basis}
\end{eqnarray}
These operators satisfy the quantum Liouville-Neumann equation
\begin{eqnarray}
i \frac{\partial}{\partial t} \hat{A} (t)
+ \Bigl[\hat{A} (t) , \hat{H} (t) \Bigr] = 0,
\nonumber\\
i \frac{\partial}{\partial t} \hat{A}^{\dagger} (t)
+ \Bigl[\hat{A}^{\dagger} (t) , \hat{H} (t) \Bigr] = 0,
\end{eqnarray}
when $u$ is a complex solution to the classical equation of motion
\begin{equation}
\ddot{u} (t) + \omega^2 (t)  u(t) = 0.
\label{cl eq}
\end{equation}
Imposing the commutation relation
\begin{equation}
\Bigl[\hat{A} (t) , \hat{A}^{\dagger} (t) \Bigr] = 1,
\end{equation}
as the annihilation and creation operators of a Fock space,
is equivalent to requiring the Wronskian  condition
\begin{equation}
m_0 \Bigl(\dot{u}^*(t) u(t) - u^* (t) \dot{u} (t) \Bigr) = i.
\label{wron}
\end{equation}

To show the equivalence between the classical equation of motion
(\ref{cl eq}) and the auxiliary equation (\ref{aux}), we
write the complex solution in a polar form
\begin{equation}
u (t) = \frac{\xi (t)}{\sqrt{2m_0}} e^{- i \theta (t)}.
\label{pol}
\end{equation}
Then Eq. (\ref{wron}) becomes
\begin{equation}
\xi^2 \dot{\theta} = 1,
\label{ang}
\end{equation}
and Eq. (\ref{cl eq}) equals to the auxiliary equation (\ref{aux}).
Furthermore, one can rewrite the operators (\ref{basis}) as
\begin{eqnarray}
\hat{A} (t) &=&  \frac{e^{-i \theta}}{\sqrt{2m_0}}
\Bigl[\frac{m_0}{\xi} \hat{q} + i (\xi \hat{p} - m_0 \dot{\xi}
\hat{q} ) \Bigr],
\nonumber\\
\hat{A}^{\dagger} (t) &=&  \frac{e^{i \theta}}{\sqrt{2m_0}}
\Bigl[\frac{m_0}{\xi} \hat{q} - i (\xi \hat{p} - m_0 \dot{\xi}
\hat{q} ) \Bigr]
\end{eqnarray}
to show that
\begin{equation}
\hat{I} (t) = \hat{A}^{\dagger} (t) \hat{A} (t) + \frac{1}{2}.
\end{equation}
The eigenstates of the invariant are the number states
\begin{equation}
\vert n, t \rangle = \frac{1}{\sqrt{n!}}
\Bigl(\hat{A}^{\dagger} (t) \Bigr)^n \vert 0, t \rangle,
\end{equation}
where the vacuum state is defined by
\begin{equation}
\hat{A} (t) \vert 0, t \rangle = 0 .
\end{equation}
Exact quantum states of the time-dependent harmonic oscillator
are given explicitly by
\begin{equation}
\vert \Psi (t) \rangle = \sum_{n} c_n \exp
\Bigl(i \int \langle n, t \vert i \frac{\partial}{\partial t}
- \hat{H} (t) \vert n, t \rangle \Bigr) \vert n, t \rangle.
\end{equation}

Second, we find one-parameter Gaussian states.
For this purpose, we choose a specific positive frequency solution
$u_0$ to Eq. (\ref{cl eq}) such that
\begin{equation}
{\rm Im} \Bigl( \frac{\dot{u}_0 (t)}{u_0(t)} \Bigr) < 0,
\end{equation}
and the Wronskian (\ref{wron}) is satisfied.
We further require that $u_0$ give the minimum uncertainty,
which will be discussed in detail in the next section.
Then any linear combination
\begin{equation}
u_{\nu} (t) = \mu u (t) + \nu^* u^* (t),
\label{lin com}
\end{equation}
also satisfies the Wronskian condition (\ref{wron}) provided that
\begin{equation}
\vert \mu \vert^2 - \vert \nu \vert^2 = 1.
\label{rel}
\end{equation}
We now make use of the complex solution (\ref{lin com})
to define
\begin{eqnarray}
\hat{A}_{\nu} (t) &=& i \Bigl( u^*_{\nu} (t) \hat{p}
- m_0 \dot{u}^*_{\nu} (t) \hat{q} \Bigr),
\nonumber\\
\hat{A}^{\dagger}_{\nu} (t) &=& -i \Bigl( u_{\nu} (t) \hat{p}
- m_0 \dot{u}_{\nu} (t) \hat{q} \Bigr).
\label{basis3}
\end{eqnarray}
Then the one-parameter Gaussian states can be found from the definition
\begin{equation}
\hat{A}_{\nu} (t) \vert 0, t \rangle_{\nu} = 0 ,
\label{vac}
\end{equation}
whose coordinate representation are given by
\begin{equation}
\Psi_{\nu} (q, t) = \Bigl(\frac{1}{2 \pi u^*_{\nu} (t)
u_{\nu} (t)} \Bigr)^{1/4}
\exp \Bigl[i \frac{m_0 \dot{u}^*_{\nu} (t) }{2 u^*_{\nu} (t)} q^2 \Bigr].
\label{Gaussian}
\end{equation}
Eq. (\ref{rel}) can be parameterized in terms
squeezing parameters  \cite{mandel}
\begin{equation}
\mu \equiv \cosh r, ~ \nu \equiv e^{i \delta} \sinh r.
\end{equation}
It follows readily that
\begin{eqnarray}
\hat{A}_{\nu} (t) &=& \tilde{\mu} \hat{A} (t)
+ \tilde{\nu} \hat{A}^{\dagger} (t),
\nonumber\\
\hat{A}_{\nu}^{\dagger} (t) &=& \tilde{\nu}^* \hat{A} (t)
+ \tilde{\mu}^* \hat{A}^{\dagger} (t),
\end{eqnarray}
where $\tilde{\mu} = \mu$, and $\tilde{\nu} = e^{i (\delta +
\pi)} \sinh r = - \nu$.
This can be rewritten as a unitary transformation
\begin{equation}
\hat{A}_{\nu} (t) = \hat{S} (z) \hat{A} (t) \hat{S}^{\dagger} (z),
\end{equation}
where
\begin{equation}
\hat{S} (z)  = \exp \Bigl[\frac{1}{2} (z^* \hat{A}_0^2
- z \hat{A}_0^{\dagger 2}) \Bigr], ~
\Bigl( z = r e^{i (\delta + \pi)} \Bigr),
\end{equation}
is a squeeze operator \cite{mandel}.
Thus one sees that $\Psi_{\nu}$ are the squeezed Gaussian states
of $\Psi_{\nu = 0}$. It should be noted that
\begin{equation}
\xi^2 (t) = 2 m_0 u_{\nu}^* (t) u_{\nu} (t),
\end{equation}
indeed satisfies the auxiliary equation (\ref{aux}).

\section{Selection Rule for Vacuum States}

In this section we study a selection rule for the vacuum states.
It will be shown that the minimum uncertainty selects uniquely
the vacuum states among the one-parameter Gaussian states in Sec. II.
For the case of time-independent harmonic oscillator
the minimum uncertainty state is the true vacuum state with the minimum
energy expectation value. For the case of time-dependent harmonic oscillator
the minimum state coincides with the Bunch-Davies vacuum state
playing a particular role in quantum field theory in a curved spacetime.
In this section we show that the one-parameter Gaussian states in Sec. II
are parameterized by the energy expectation value and
are the squeezed states of the true vacuum state.
For time-dependent case we prove an inequality between
the energy expectation value of a squeezed Gaussian state
and that of the minimal squeezed Gaussian state.

\subsection{Time-Independent Case: True Vacuum}

For the case of a time-independent harmonic oscillator
we can show that the squeezing parameter of one-parameter Gaussian
states is nothing but the energy expectation value.
The energy expectation value of the Hamiltonian with
respect to the Gaussian state (\ref{Gaussian}) is given by
\begin{equation}
{}_{\nu}\langle 0, t \vert \hat{H} \vert 0, t \rangle_{\nu} =
\frac{1}{4} \Bigl(\dot{\xi}^2 + \omega^2 \xi^2 + \frac{1}{{\xi}^2} \Bigr)
\equiv \epsilon.
\label{ham ex}
\end{equation}
Equation (\ref{aux}) can be integrated to yield Eq. (\ref{ham
ex}). We solve the integral equation (\ref{ham ex}) to obtain
\begin{equation}
\xi^2 = \frac{2 \epsilon}{\omega^2} + \frac{2 \epsilon}{\omega^2}
\sqrt{1 - \frac{\omega^2}{4 {\epsilon^2}}} \cos (2 \omega t).
\label{xi sq}
\end{equation}
By solving (\ref{ang}) we get
\begin{equation}
\theta = \omega t.
\end{equation}

We now compare the $\xi^2$ of Eq. (\ref{xi sq})
with that obtained by solving directly Eq. (\ref{cl eq}).
We choose the following specific solution to Eq. (\ref{cl eq})
\begin{equation}
u_0 (t) = \frac{1}{\sqrt{ 2 m_0 \omega}} e^{- i \omega t},
\label{sp sol}
\end{equation}
and confine our attention to  real $\mu$ and $\nu$.
It then follows that
\begin{equation}
\xi^2 = \frac{1}{\omega} \Bigl(\mu^2 + \nu^2
+ 2 \mu \nu \cos(2 \omega t) \Bigr).
\label{xi sq2}
\end{equation}
By comparing Eqs. (\ref{xi sq}) and (\ref{xi sq2}) we find the
squeezing parameter
\begin{eqnarray}
\mu = \sqrt{\frac{\epsilon}{\omega} + \frac{1}{2}},
\nonumber\\
\nu = \sqrt{\frac{\epsilon}{\omega} - \frac{1}{2}}.
\end{eqnarray}
Thus we were able to express the squeezing parameters
in terms of the energy expectation value.

We now look for the Gaussian state with the minimum uncertainty.
The one-parameter Gaussian states have the uncertainty
\begin{equation}
(\Delta p)_{\nu} (\Delta q)_{\nu} = \frac{1}{2} \Bigl(
\vert \mu \vert^2 + \vert \nu \vert^2 \Bigr).
\end{equation}
The minimal uncertainty state is obtained by $\mu = 1$ and $\nu = 1$
and the uncertainty is $1/2$.
This has also the minimum energy
\begin{equation}
\epsilon_{\rm min.} = \frac{\omega}{2}.
\end{equation}
So the specific solution (\ref{sp sol}) corresponds to the
minimum energy and the corresponding Gaussian state is the true
vacuum state of the harmonic oscillator.
Therefore, the Gaussian states we have found in Sec. II
are the one-parameter squeezed states of the true vacuum
state whose parameter is the energy expectation value.

\subsection{Time-Dependent Case: Bunch-Davies Vacuum}

We now turn to the time-dependent case.
In general, one can show the following inequality of
the uncertainty relations with respect to $\Psi_{\nu}$ and $\Psi_{0}$
\begin{eqnarray}
(\Delta p)_{\nu} (\Delta q)_{\nu} &=& m_0
\Bigl(\dot{u}_{\nu}^* (t) \dot{u}_{\nu} (t)
u_{\nu}^* (t) u_{\nu} (t) \Bigr)^{1/2}
\nonumber\\
&\geq& m_0 \Bigl( \mu - \vert \nu \vert \Bigr)^2
\Bigl(\dot{u}_{\nu = 0}^* (t) \dot{u}_{\nu = 0} (t)
u_{\nu = 0}^* (t) u_{\nu = 0} (t) \Bigr)^{1/2}
\nonumber\\
&\geq& m_0 \Bigl(\dot{u}_{\nu = 0}^* (t) \dot{u}_{\nu = 0} (t)
u_{\nu = 0}^* (t) u_{\nu = 0} (t) \Bigr)^{1/2}
\nonumber\\
&=& (\Delta p)_{\nu = 0} (\Delta q)_{\nu = 0}.
\label{un ineq}
\end{eqnarray}
The equality of Eq. (\ref{un ineq}) holds when $\mu = 1$ and $\nu
= 0$. The energy expectation value similarly satisfies the inequality
\begin{eqnarray}
\langle \Psi_{\nu} \vert \hat{H} \vert \Psi_{\nu} \rangle &=&
m_0 \Bigl(\dot{u}_{\nu}^* (t) \dot{u}_{\nu} (t)
+ \omega^2 (t) u_{\nu}^* (t) u_{\nu} (t) \Bigr)
\nonumber\\
&\geq&
\Bigl( \mu - \vert \nu \vert \Bigr)^2
m_0 \Bigl(\dot{u}_{\nu = 0}^* (t) \dot{u}_{\nu = 0} (t)
+ \omega^2 (t) u_{\nu = 0}^* (t) u_{\nu = 0} (t) \Bigr)
\nonumber\\
&\geq&
m_0 \Bigl(\dot{u}_{\nu = 0}^* (t) \dot{u}_{\nu = 0} (t)
+ \omega^2 (t) u_{\nu = 0}^* (t) u_{\nu = 0} (t) \Bigr)
\nonumber\\
&=&
\langle \Psi_{\nu = 0} \vert \hat{H} \vert \Psi_{\nu = 0} \rangle,
\label{en ineq}
\end{eqnarray}
where the equality holds when $\mu = 1$ and $\nu = 0$.
What Eqs. (\ref{en ineq}) and (\ref{un ineq}) implie is that
once we choose the Gaussian state with the minimum uncertainty and
energy at each moment, all its squeezed Gaussian states have
higher uncertainty and energy. However, it should be reminded that
the energy expectation value for a time-dependent quantum system
does not have an absolute physical meaning since it is not
conserved. On the other hand, the quantum uncertainty still has
some physical meanings even for the time-dependent quantum system
in that it characterizes the very nature of quantum states.
For this reason, we put forth the {\it minimum uncertainty} as the selection
rule for the vacuum state for time-dependent system. From Eqs.
(\ref{en ineq}) and (\ref{un ineq}), the vacuum state with the
minimum uncertainty has also the minimum energy at every moment.

In order to show that the vacuum state with the minimum
uncertainty coincides indeed with the well-known vacuum states
we consider a minimal massive scalar field in the de Sitter
spacetime. The de Sitter spacetime has the metric
\begin{equation}
ds^2 = - dt^2 + e^{2 H_0} d{\bf x}^2,
\end{equation}
where $H_0$ is an expansion rate of the universe.
When the massive scalar field is decomposed into
Fourier modes, it has the Hamiltonian
\begin{equation}
H = \sum_{{\bf k}, (\pm)} \frac{e^{-3 H_0 t}}{2} (\pi_{\bf
k}^{(\pm)})^2 + \frac{e^{3 H_0 t}}{2} (m^2 + {\bf k}^2 e^{- 2 H_0 t}
) (\phi_{\bf k}^{(\pm)})^2,
\end{equation}
where $u_{\bf k}^{(\pm)}$ denote the cosine- and sine-modes, respectively.
Thus the massive scalar field system is equivalent to infinitely many
harmonic oscillators both with a time-dependent mass and with
time-dependent frequencies. Though the mass depends on time,
all the previous results are valid only with the following modification
\begin{eqnarray}
\hat{A}_{\bf k}^{(\pm)} &=& i \Bigl( u_{\bf k}^{(\pm) *} (t)
\hat{\pi}_{\bf k}^{(\pm)} - e^{3 H_0 t} \dot{u}_{\bf k}^{(\pm) *} (t)
\hat{\phi}_{\bf k}^{(\pm) *} \Bigr),
\nonumber\\
\hat{A}_{\bf k}^{(\pm) \dagger} &=& - i \Bigl( u_{\bf k}^{(\pm)} (t)
\hat{\pi}_{\bf k}^{(\pm)} - e^{3 H_0 t} \dot{u}_{\bf k}^{(\pm)} (t)
\hat{\phi}_{\bf k}^{(\pm)} \Bigr),
\end{eqnarray}
where $u_{\bf k}^{(\pm)} (t)$ satisfy the equations
\begin{equation}
\ddot{u}_{\bf k}^{(\pm)} (t) + 3 H_0 u_{\bf k}^{(\pm)} (t)
+ (m^2 + {\bf k}^2 e^{- 2 H_0 t}) u_{\bf k}^{(\pm)} (t) = 0.
\end{equation}
It can be shown \cite{kim3} that the specific solution in the Hankel function
of the second kind
\begin{equation}
u_{\bf k}^{(\pm)} (t) = \Bigl( \frac{\pi}{4 H_0} \Bigr)^{1/2} e^{- \frac{3}{2} H_0 t}
H_{\chi}^{(2)} (z),
\label{hankel}
\end{equation}
where
\begin{equation}
\chi = \Bigl(\frac{9}{4} - \frac{m^2}{H_0^2} \Bigr)^{1/2},~
z = \frac{k}{H_0} e^{- H_0 t},
\end{equation}
gives rise to the Gaussian state with the minimum uncertainty at each moment.
Moreover, the Gaussian state has the uncertainty $1/2$
at earlier times $t \rightarrow - \infty$. The vacuum
state of the scalar field
\begin{equation}
\vert 0, t \rangle_{\nu = 0} = \prod_{{\bf k}, \pm}
\vert 0_{\bf k}^{(\pm)} \rangle_{\nu = 0}
\end{equation}
is indeed the Bunch-Davies vacuum state \cite{bunch}.
The general solution of the form (\ref{lin com}) mixes
the positive frequency solution $u_{\bf k}^{(\pm)}$
with the negative frequency solution $u_{\bf k}^{(\pm) *}$.

\section{Conclusion}

In this paper we have found the one-parameter squeezed Gaussian states
for a time-dependent harmonic oscillator. It was found that the
squeezing parameters can be expressed in terms of energy
expectation value and represents the degree of mixing of positive
and negative frequency solutions. The {\it minimum uncertainty} is
advanced as a selection rule for the vacuum state. We have illustrated
the selection rule for the vacuum state by studying a time-independent
harmonic oscillator and a minimal massive scalar field in a de Sitter spacetime.
It was shown that the Gaussian states with the minimum uncertainty are
the true vacuum state with the minimum energy for the time-independent
harmonic oscillator and the Bunch-Davies vacuum state for the
massive scalar field in the de Sitter spacetime.

\acknowledgments

We are deeply indebted to J. H. Yee for many valuable
discussions and comments. We also thank J. Aliaga and
V. I. Man'ko for informing useful references.
This work was supported by the Non-Directed Research Fund, Korea
Research Foundation, 1997.

\end{document}